\documentclass{pasa}

\usepackage{amsmath}
\usepackage{amssymb}
\usepackage{times}
\usepackage{pifont} 

\newcommand{\teff}{$T_{\mathrm{eff}}$}
\newcommand{\muhz}{$\mu$Hz}
\newcommand{\numax}{$\nu_{\mathrm{max}}$}
\newcommand{\dnu}{$\Delta\nu$}

\newcommand{\kepler}{\textit{Kepler}}

\newcommand{\msol}{M$_\odot$}

\title[Suppression of quadrupole and octupole modes]{Suppression of quadrupole and octupole modes in red giants observed by \textit{Kepler}}
\author[Stello et al.]
{Dennis~Stello$^{1,2,3}$, Matteo~Cantiello$^4$,
 Jim~Fuller$^{5,4}$, Rafael A. Garcia$^{6}$, \and Daniel~Huber$^{1,2}$\\
\affil{$^1$Sydney Institute for Astronomy (SIfA), School of Physics, University of Sydney, NSW 2006, Australia}%
\affil{$^2$Stellar Astrophysics Centre, Department of Physics and
  Astronomy, Aarhus University, Ny Munkegade 120, DK-8000 Aarhus C,
  Denmark}
\affil{$^3$School of Physics, University of New South Wales, NSW 2052, Australia}
\affil{$^4$Kavli Institute for Theoretical Physics, University of California, Santa Barbara, CA 93106}
\affil{$^5$TAPIR, Walter Burke Institute for Theoretical Physics, Mailcode 350-17 California Institute of Technology, Pasadena, CA 91125}
\affil{$^6$Laboratoire AIM, CEA/DSM -- CNRS -- Univ. Paris Diderot -- IRFU/SAp Centre de Saclay, 91191 Gif-sur-Yvette Cedex, France}
}%
\jid{PASA}
\doi{10.1017/pas.\the\year.xxx}
\jyear{\the\year}

\usepackage[authoryear]{natbib}
\bibpunct{(}{)}{;}{a}{}{,}
\usepackage{aas_macros}
\usepackage{hyperref} 
\hypersetup{colorlinks,citecolor=blue,linkcolor=blue,urlcolor=blue}

\begin{document}

\begin{abstract}
The asteroseismology of red giant stars has continued to yield surprises since the
onset of high-precision photometry from space-based observations.
An exciting new theoretical result shows that the previously observed
suppression of dipole oscillation modes in red giants can be used to detect
strong magnetic fields in the stellar cores. 
A fundamental facet of the theory is that nearly all the mode energy
leaking into the core is trapped by the magnetic greenhouse effect.  This
results in clear predictions for how the mode visibility changes 
as a star evolves up the red giant branch, and how that depends on
stellar mass, spherical degree, and mode lifetime.  Here, we 
investigate the validity of these predictions with a
focus on the visibility of different spherical degrees.  We find
that mode suppression weakens for higher degree modes with an average 
reduction in the quadrupole mode visibility of up to 49\% for the least
evolved stars in our sample, 
and no detectable suppression of octupole modes, in agreement with the
theoretical predictions. We furthermore find evidence for the influence of
increasing mode lifetimes on the measured visibilities along the red giant
branch, in agreement with previous independent observations.  These
results support the theory that strong internal magnetic fields are
responsible for the observed suppression of non-radial modes in red giants.
We also find preliminary evidence that stars with suppressed dipole modes
on average have slightly lower metallicity than normal stars.
\end{abstract}


\maketitle%


\section{Introduction} 
The asteroseismology of red giant stars has become a highlight of
the CoRoT and \kepler\ space missions \citep[for general reviews see
e.g.][]{ChaplinMiglio13,GarciaStello15}.  
One feature that 
has made these stars interesting, is the presence of non-radial
oscillation modes that reveal properties of the stellar cores.  The
non-radial modes have a mixed nature, behaving like acoustic (or p) modes
in the envelope with 
pressure acting as the restoring force,
and as gravity (or g) modes in the core region with buoyancy being the
restoring force \citep[e.g.][]{BeddingTenerife}.  The p- and g-mode cavities are
separated by an evanescent 
zone, which the waves can tunnel through from either side. 
The exact p- and g-`mixture', or flavor, of a mixed mode
depends on its frequency and spherical degree, $\ell$.  Modes with
frequencies close to the acoustic resonant frequencies of the stellar
envelope tend to be more p-mode like, while those far from the acoustic
resonances are much more g-mode like.  The latter therefore probe deeper
into the stellar interior compared to the former.  How much the flavor changes from
mode to mode across the acoustic resonances depends on the overall coupling 
between the envelope and the core.  The overall aspects of mode mixing in
red giants arising from this coupling 
is well understood theoretically \citep{Dupret09}.   
Observationally, the dipole modes ($\ell=1$) have turned out to be
particularly useful probes of the 
core because of their stronger
coupling between core and envelope.  The characterization of dipole mixed modes
\citep{Beck11} led to the discovery that red
giant branch stars can be clearly distinguished from red clump stars
\citep{Bedding11}, and to the detection of radial differential rotation
\citep{Beck12}. 
Modes of higher spherical degree are also mixed, but the weaker coupling
makes it difficult to observe the modes with strong g-mode
flavor. The quadrupole modes ($\ell=2$) we observe, and to a larger degree the
octupole modes ($\ell=3$), are therefore on average more acoustic compared to the
dipole modes, and hence less sensitive to the stellar core. 

One particular observational result about the dipole modes posed an
intriguing puzzle. The ensemble study of a few hundred \kepler\ red giants by 
\citet{Mosser12a} showed that a few dozen stars -- about
20\% of their sample -- had significantly lower power in the dipole
modes than `normal' stars.  However, no significant suppression of higher
degree modes was reported, leading to the conclusion only dipole modes
were affected.  

Recent theoretical work has proposed that the mechanism responsible for the
mode suppression results in almost total trapping of the mode energy
that leaks into the g-mode cavity \citep{Fuller15}. They put forward
magnetic fields in the core region of the stars as the most plausible
candidate for the suppression mechanism. 
This interpretation was further supported by the observation that mode
suppression only occurs in stars above 1.1\msol, with an increasing
fraction up to 50-60\% for slightly more massive stars, all of which hosted
convective cores during their main sequence phase; strongly pointing to a
convective core dynamo as the source of the mode suppressing magnetic field 
\citep{Stello16}.  
Both \citet{Fuller15} and \citet{Stello16} focused 
their analysis on 
dipole modes. However, the theory by \citet{Fuller15} does 
allow one to predict the magnitude of the suppression for higher degree
modes. Agreement with observations of these modes would provide important
support for the proposed mechanism.

In this paper, we use 3.5 years of \kepler\ data of over
3,600 carefully selected red giant branch stars to investigate the mode
suppression in the non-radial modes of degree $\ell=1$, 2, and 3,
and compare theoretical predictions with our observed mode visibilities.

\section{Theoretical predictions} \label{theory}
We use the Modules for Experiments in Stellar Evolution 
\citep[MESA, release \#7456,][]{Paxton11,Paxton13,Paxton15} 
to compute stellar evolution
models 
of low-mass stars from the zero age main sequence to the tip of 
the red giant branch. Non-rotating models have been computed using an
initial metallicity of $Z=0.02$ with a mixture taken from
\citet{Asplund05} and adopting the OPAL opacity tables
\citep{IglesiasRogers96}. We calculate convective
regions using the mixing-length theory with $\alpha_{\rm MLT}=2.0$. The
boundaries of convective regions are determined according to the
Schwarzschild criterion. 

We calculate the expected visibilities for dipole, quadrupole, and octupole
modes in models with 
masses $1.1\,$\msol,  $1.3\,$\msol,
$1.5\,$\msol,  $1.7\,$\msol, and $1.9\,$\msol\ following
\citet{Fuller15}. According to the theory, the ratio of suppressed mode
power to normal mode power is  
\begin{equation}
\label{eqn:vsup}
\frac{V_{\rm sup}^2}{V_{\rm norm}^2} = \bigg[ 1 + \Delta \nu\, \tau \,T^2 \bigg]^{-1} \, ,
\end{equation}
where \dnu\ is the large frequency separation, $\tau$ is the radial
mode lifetime measurable from the observed frequency power spectrum
\citep[e.g.][]{Corsaro15}), and $T$ is the wave transmission coefficient
through the evanescent zone. $T$ is calculated via  
\begin{equation}
\label{eqn:T}
T  = \exp \bigg[ - \int^{r_2}_{r_1} dr \sqrt{ -\frac{ \big( L_\ell^2 - \omega^2 \big) \big(N^2 - \omega^2 \big) }{v_s^2 \omega^2} } \bigg] \, .
\end{equation}
Here, $r_1$ and $r_2$ are the lower and upper boundaries of the evanescent
zone, $L_\ell^2 = l(l+1)v_s^2/r^2$ is the Lamb frequency squared, $N$ is
the buoyancy frequency, $\omega$ is the angular wave frequency, and
$v_s$ is the sound speed. We calculate \dnu\ and the frequency of
maximum power, \numax, using the scaling relations of
\cite{Brown91} and \citet{KjeldsenBedding95} with the solar references
values, \dnu$_\odot=135.1\,$\muhz\ and \numax$_\odot=3090\,$\muhz.

\section{Data analysis}

\subsection{Sample selection}\label{sampleselection}
Our initial selection of stars was based on the analysis by
\citet{Stello13} of 13,412 red giants from which we adopt their
measurements of \dnu\ and \numax, as well as their estimates of stellar
mass based on the combination of those seismic observables with
photometrically-derived effective 
temperatures using scaling relations \citep[see][~for details]{Stello13}. 
In order to select only red giant branch stars from this
sample we follow the approach by \citet{Stello16}. 
We selected all 3,993 stars with \numax\ $> 50\,$\muhz\ and $M<
2.1\,$\msol, which based on stellar models 
is expected to select only red giant branch stars
\citep[e.g.][]{Stello13}.  However, this selection 
do not take measurement
uncertainties in \numax\ and $M$ into account, possibly introducing some
helium-core burning stars into our sample towards the lower end of the
\numax\ bracket.   
We therefore performed a few additional steps to further reduce the
possible `contamination' by helium-burning stars in our sample.

\begin{figure}
\includegraphics[width=8.5cm]{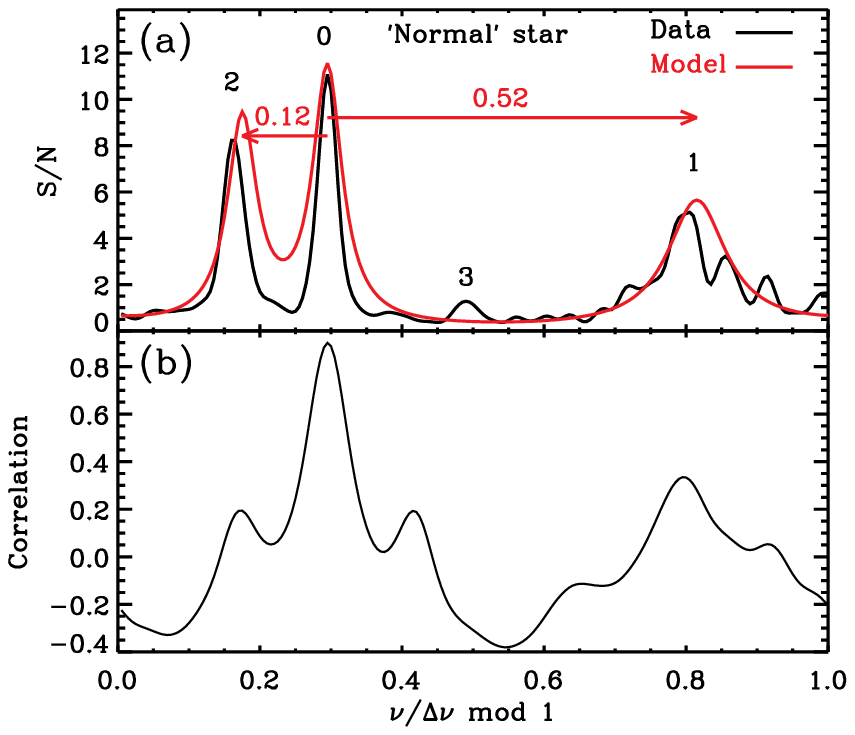}
\includegraphics[width=8.5cm]{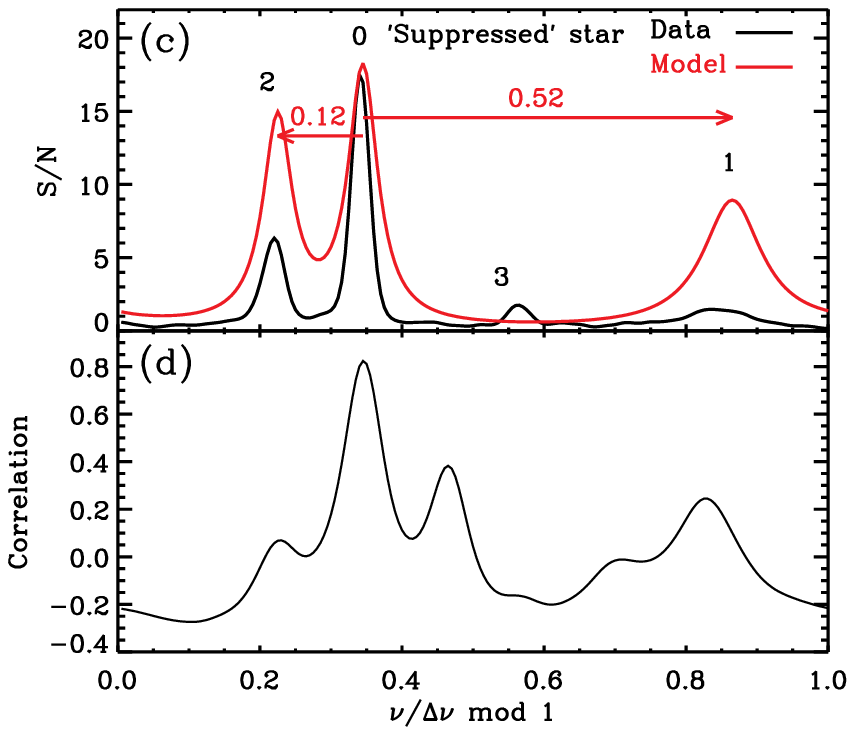}
\caption{(a) Folded and smoothed spectrum (black
  curve) of the central 4\dnu-wide region around \numax\ of a normal star
  (KIC2141255).  Regions dominated by modes of degrees, $\ell=0$, 1, 2, and 3
  are indicated. The red curve shows the model shifted to the
  position resulting in the largest correlation with the data. (b)
  Correlation versus shift between model and data.  
  (c) and (d) The same as
  (a) and (b), but for a star with suppressed non-radial modes (KIC4348666).  
\label{foldedspectrum}} 
\end{figure}

\begin{figure}
\includegraphics[width=8.8cm]{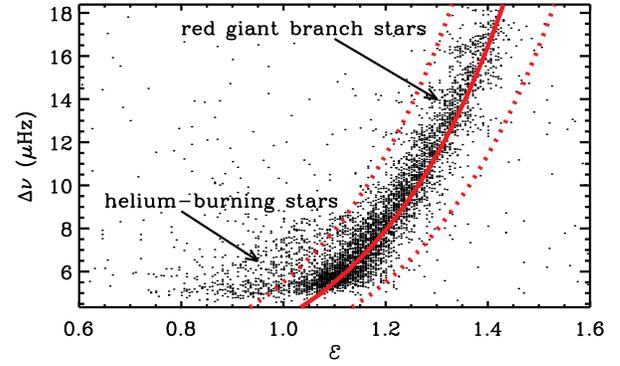}
\caption{\dnu\ versus $\varepsilon$ of the 3,993 red giants from the
  \citet{Stello13} sample with \numax\ $> 50\,$\muhz\ and $M< 2.1\,$\msol. 
  The solid red curve shows the relation by \citet{Corsaro12}, and the
  dotted red curves show the relation shifted by $\pm 0.1$. 
\label{epsilon}} 
\end{figure}

First, we derived the offset from zero of the harmonic series of radial
modes, known as $\varepsilon$ in the asymptotic relation, 
$\nu=\Delta\nu(n+\ell/2+\varepsilon)$, where $\nu$ is
the mode frequency and $n$ is the radial order.
This offset is known to 
be different for helium-burning
stars \citep{Kallinger12,Dalsgaard14} relative to red giant branch stars of
similar density, and hence has diagnostic power to distinguish the two
types of stars.
To measure $\varepsilon$ we calculated frequency power
spectra for each star using long-cadence ($\Delta t \simeq 29.4\,$minutes)
\kepler\ data obtained between 2009 May 2 and 2012 Oct 3, corresponding to
observing quarters 0--14, or a total of about 54,000 data
points per star.  
We used the method by \citet{Huber09} to fit and remove the background
noise profile of the spectra, and then selected the central $\pm2$\dnu\
spectral range around \numax, which we folded with an interval of \dnu.
The folded spectrum was smoothed by a Gaussian function with a
full-width-half-maximum of 0.1\dnu, and correlated with a 
model of the spectrum (Figure~\ref{foldedspectrum}). 
Each spherical degree in the model was described by a Lorentzian
profile with relative heights of 1.0, 0.5, and 0.8 for the $\ell=0$,
$\ell=1$, and 
$\ell=2$ modes, and widths of 5\%, 10\%, and 5\% of \dnu, respectively.  The
Lorentzian profiles were centered relative to each other with the one
representing quadrupole modes being 0.12\dnu\ to the left of the radial
mode profile, and the one representing the dipole modes being 0.52\dnu\ to
the right of the radial mode following the results by \citet{Huber10} (see
Figure~\ref{foldedspectrum}).  The shift  
between data and model providing the largest correlation was adopted as
 $\varepsilon\!\! \mod 1$.  
Only the 3,721 stars with $\varepsilon$ within $\pm0.1$ of the \dnu-$\varepsilon$
relation by \citet{Corsaro12} ($\varepsilon = 0.634+0.63\log(\Delta\nu)$)
were kept in our sample 
(see Figure~\ref{epsilon}).  

Second, we cross matched our remaining sample with the stars
identified as burning helium by \citet{Stello13} and \citet{Mosser14}
based on their dipole-mode period spacings.  We found
86 helium-burning stars this way, which we 
removed.  We note
that all but eight of these helium-burning 
contaminants had \numax\
$< 70\,$\muhz\ (and all $<85\,$\muhz).  Although our remaining sample  
could potentially still include some helium-burning stars if they were not
in the \citet{Stello13} and \citet{Mosser14} samples, the above
cross match indicates such contaminants would most likely have \numax\ $<
70\,$\muhz.  
As a final check, we visually inspected the power spectra of the remaining
sample (3,635 stars), which led to the removal of 
24 stars that
appeared to be helium-core burning or had spectra of such poor quality (low
signal-to-noise or bad window function) that the stellar classification was
ambiguous.  

\subsection{Mode visibility measurements}
\begin{figure}
\includegraphics[width=8.5cm]{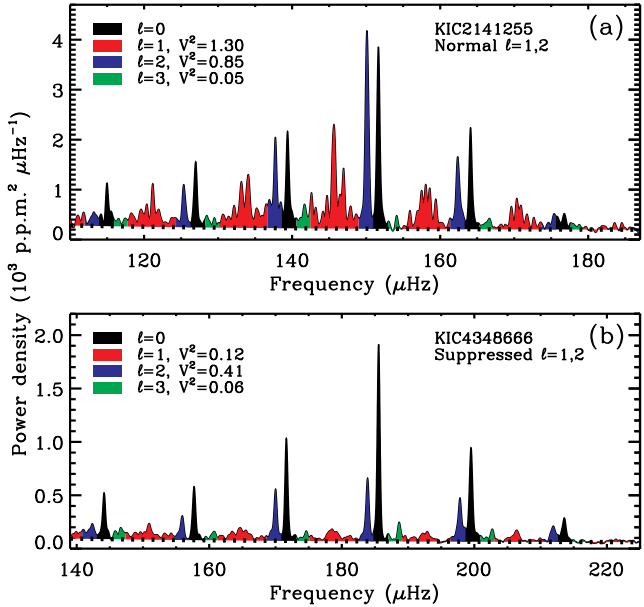}
\caption{Representative power spectra of two red giants from our
  sample. For clarity the 
  spectra have been smoothed by 0.03\dnu, which partly merges the power from
  individual mixed modes into one broad peak around each acoustic
  resonance. The colored regions indicate the integrated power associated
  with each mode of degree $\ell=0$, 1, 2, and 3. The horizontal dashed
  curves shows the estimated noise profile (a) Star with normal mode
  visibility, $V^2$, of the non-radial modes. (b) Star with low mode
  visibilities in both dipole and quadrupole modes, but normal octupole
  mode visibility.  
\label{examplespectra}} 
\end{figure}

To measure the mode visibilities, we first needed to identify the 
regions in the power spectra dominated by the different modes.
Here, we used the $\varepsilon$ values found in the previous step
(Sect.~\ref{sampleselection}) to locate the four central radial modes
closest to \numax, and followed the approach by \citet{Stello16} for masking
the parts of the spectra dominated by $\ell=0$, $\ell=1$, $\ell=2$, and
$\ell=3$ modes. The regions we chose were 
$\varepsilon-0.06 < (\nu/\Delta\nu \mod 1) < \varepsilon+0.10$ for $\ell=0$,
$\varepsilon+0.25 < (\nu/\Delta\nu \mod 1) < \varepsilon+0.78$ for $\ell=1$,
$\varepsilon-0.22 < (\nu/\Delta\nu \mod 1) < \varepsilon-0.06$ for $\ell=2$, and 
$\varepsilon+0.10 < (\nu/\Delta\nu \mod 1) < \varepsilon+0.25$ for $\ell=3$.
In this way, the entire 4\dnu-wide central part of the spectrum was divided
up into distinct segments associated with either $\ell=0$, $\ell=1$,
$\ell=2$, or $\ell=3$ modes. In Figure~\ref{examplespectra} we show a couple
of representative spectra with the results of the masking, identifying the
different mode degrees.  
See also Figure 1 in \citet{Stello16} who used the
exact same scheme. 
Finally, the mode visibilities, $V^2$, of the non-radial modes were derived
as the ratio of the total mode power of the segments associated with a
given degree relative to the total radial mode power.  For this we used the
background corrected spectra described in Sect.~\ref{sampleselection}.  
We note that the uncertainty in the background estimation introduces
measurement scatter on top of the intrinsic spread in the visibilities,
potentially pushing intrinsically low values of $V^2$ below zero.
Rejecting those stars or artificially pushing them up to $V^2=0$ would be
statistically incorrect and bias our measurements of the average $V^2$.

\section{Results}

\subsection{The dipole modes revisited}
\begin{figure}
\includegraphics[width=8.5cm]{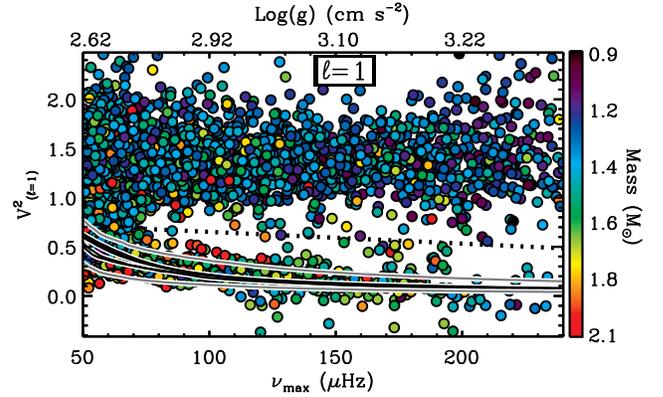}
\caption{Visibility of dipole modes ($V^2_{\ell=1}$) for 3,611 red giants
  below the red giant branch luminosity bump. Stars evolve from right to
  left, and their approximate $\log g$ values are shown on the top
  axis.  The rms scatter in $V^2_{\ell=1}$ is 0.26 (or 20\%) for the normal
  stars. 
  The color of each symbol indicates the stellar mass according to the
  scale on the right-hand side.
  The solid black lines are the predicted visibilities from 1.1\msol, 1.3\msol,
  1.5\msol, 1.7\msol, and 1.9\msol\ models 
  adopting an average mode lifetime 
  of 20 days in agreement with \citet{Corsaro15} (Sect,~\ref{theory}).
  The solid gray curves are the predicted visibilities from 1.7\msol\ models
  for mode lifetimes of 10 days (highest $V^2_{\ell=1}$) and 40 days (lowest
  $V^2_{\ell=1}$).  
  The dotted fiducial line, adopted from \citet{Stello16}, separates
  dipole-suppressed and normal stars. The observed dipole visibilities are
  identical to those by \citet{Stello16}.
\label{vis_vs_numax}} 
\end{figure}

In Figure~\ref{vis_vs_numax} we show our measurements of the dipole mode
visibility as function of \numax.  
We denote the stars below the dotted line as the dipole-suppressed sample,
as opposed to the `normal' stars.  
As shown in \citet{Fuller15} and
\citet{Stello16}, the predicted visibilities of the suppressed dipole modes
(solid lines) match the observations remarkably well, and with normal stars
on average being less massive.  Here, we show that the
prediction for the dipole-mode suppression is insensitive to
stellar mass for the typical mass range of the \kepler\ red giants (the
solid black curves fall almost on top of each other), which is also what we
observe in the data.  The slight dependence on mode lifetime, $\tau$, is
illustrated by comparing the black curves (all $\tau=20\,$days) with the
gray curves based on a $1.7\,$\msol\ track just above ($\tau=10\,$days) and
below ($\tau=40\,$days) the black curves.  
We find no variation in the average visibility ($V^2_{\ell=1}=1.35$) and
its scatter for the normal stars as a function on \numax.   

In an attempt to determine if the dipole-suppressed sample is somehow
distinct other than by their mass \citep{Stello16}, we do not find them to be
statistically different in terms of their distance and position on sky.
If we focus on the mass range $1.5 < M$/\msol\ $<1.8$, where we find
the highest number and fraction (50--60\%) of dipole-suppressed stars
\citep{Stello16}, the dipole-suppressed sample is on 
average slightly more metal poor (by $0.045\pm0.017$ dex) than the normal stars,
and slightly hotter (by  $26.9\pm8.9\,$K) \citep[based on SDSS-APOGEE
DR12,][]{Alam15}.  

\subsection{The quadrupole and octupole modes}
\begin{figure}
 \includegraphics[width=8.5cm]{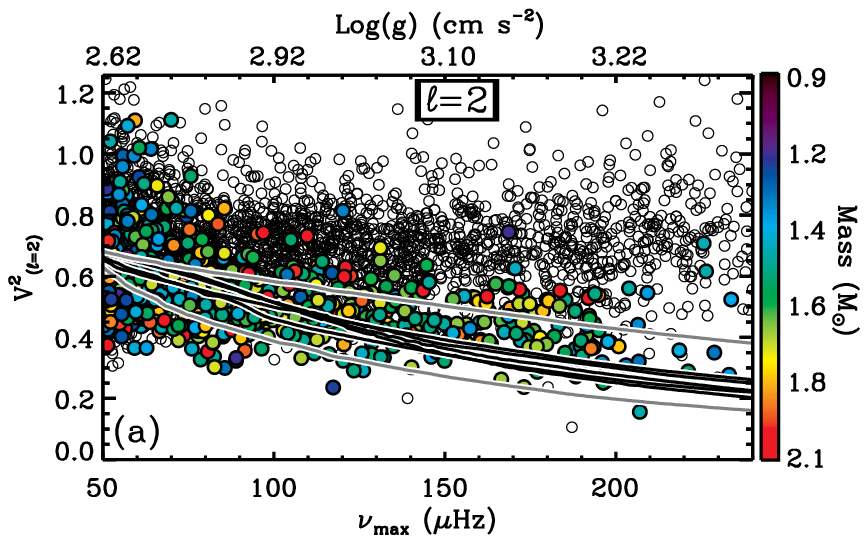}
 \includegraphics[width=8.5cm]{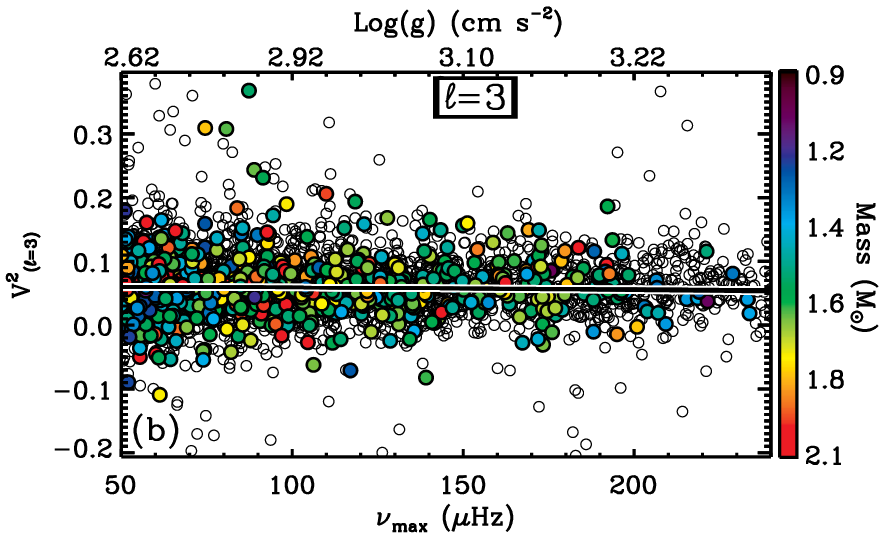}
 \caption{(a) Visibility of quadrupole modes for the same stars as shown in
  Figure~\ref{vis_vs_numax}.  The notation follows that of
  Figure~\ref{vis_vs_numax} except only the dipole-suppressed stars (those below
  the dotted line in Figure~\ref{vis_vs_numax}) are shown by filled symbols. 
  The observed rms scatter in $V^2_{\ell=2}$ is 0.14 (or 20\%) for
  the normal stars. 
  (b) The same as panel (a) but for octupole modes.  Here, the
  observed $V^2_{\ell=3}$ scatter is 0.05 (84\%). 
\label{vis2_vs_numax}}
\end{figure}

The quadrupole and octupole visibilities are shown in
Figures~\ref{vis2_vs_numax} (a) and (b), respectively. Here, we only show
the dipole-suppressed sample as filled symbols, to make it easier to
distinguish them from the normal population shown as empty symbols.
While less dramatic than the dipole modes, the quadrupole modes show
significant reduction in mode power for the dipole-suppressed
sample.  The least evolved stars ($210 < \,\,$\numax/\muhz\ $< 240$)
show on average $V^2_{\ell=2}=0.350\pm 0.053$, compared to
$V^2_{\ell=2}=0.688\pm 0.003$ for all the normal stars, corresponding to 
49\% reduction in power. 
Although there is overall good agreement between the predictions and the 
observations, the trend of the visibility as a function of 
\numax\ is predicted to be steeper than observed if we, as done here,
assume a fixed mode lifetime along the red giant branch 
(Figure~\ref{vis2_vs_numax}(a)).  We used a lifetime of $20\,$days, which is
representative for the average value found by \citet{Corsaro15}.  This 
assumption could, at least partly, explain the discrepancy,
because the observed mode lifetimes are most likely increasing as red giants
evolve towards lower \numax\ and lower \teff.  The variation in
mode lifetime was found to be roughly 50\% by \citet[][their
Figure 7]{Corsaro15} across a sample spanning a \numax\ range of $\sim
110-160\,$\muhz\ (their Table A.2).  Our sample span a much larger \numax\
range, and for reference we therefore also show the $1.7\,$\msol\ tracks 
in gray with factors of two difference in mode lifetime; 10 and 40
days, respectively.  Based on this, we conclude that an increasing mode
lifetime along the red giant branch could possibly explain the apparent
discrepancy of the predicted versus observed trend seen in
Figure~\ref{vis2_vs_numax}(a).  A mode lifetime of $\sim 15\,$days at
\numax\ $= 200\,$\muhz\ and a lifetime of $\sim 30\,$days at \numax\ $=
70\,$\muhz\ would provide a better match to the data. 
However, we note that the predicted visibilities of the suppressed sample
depends somewhat on the overshoot beyond classical Schwarzschild implemented
in the stellar models (Cantiello et al. 2016, in prep.), so we defer a more
thorough analysis to future work, but point out here the possibility of
constraining either the mode lifetime or the overshoot given independent
measurements of the other.    

Turning to the octupole modes (Figure~\ref{vis2_vs_numax}(b)), we see no
significant effect of the suppression, as anticipated by theory (solid
black curves) because they are almost purely acoustic envelope modes.  We
find the average visibilities to be $V^2_{\ell=3}=0.0591\pm0.0009$ for the
normal stars, and $V^2_{\ell=3}=0.0577\pm0.0018$ for the dipole-suppressed sample,
which are the same within $1\sigma$.  In Figure~\ref{vis2_vs_numax}(b) we do
not show the tracks of different mode lifetimes because they are
essentially indistinguishable from the black curves.

\subsection{Discussion}
Both the evolution of the predicted mode visibility as a function of
\numax\ (for fixed spherical degree) and its dependence on mode degree (for
fixed \numax) are essentially determined by the difference in
coupling between the core and envelope as the star evolves (or as seen by
modes of different degree).  The key aspect of this predicted behavior is
that there is perfect trapping of the mode energy leaking into the core
by the suppression mechanism.  Hence, given the agreement reported here
between theory and observations for both $\ell=2$ and $\ell=3$ modes, in
addition to what is seen for dipole modes, supports this notion that the
mode suppression mechanism indeed traps almost, if not, all the energy that
leaks into the core region. 

The theoretical predictions presented in
Figures~\ref{vis_vs_numax} and ~\ref{vis2_vs_numax} are derived
relative to non-suppressed stars, and are hence scaled to the observed
average visibilities of the normal stars.
We note that the absolute scale of the observed visibilities reported here 
are not directly comparable with expectations for the normal stars
\citep{Ballot11} because we simply integrated the power over fixed 
regions of the spectrum in terms of \dnu.  Due to the presence of mixed
modes, which strictly speaking exist across the entire spectrum, small
amounts of power (from gravity-dominated modes far from the acoustic
resonant frequencies) from one degree of modes will
be present in the regions that we associate with modes of another degree.    
Direct comparisons would therefore need to be based on simultaneous fitting 
of model profiles to all the modes in the spectrum,
which 
is currently not practically possible 
for large numbers of stars
\citep{Corsaro15}.  Similarly, any differences between the average
visibilities reported here and those by \citet{Mosser12a}, are most likely
due to slight different choices for the mode integration regions in the two
studies. 



\section{Summary and outlook}
We have measured oscillation mode power in over 3,600 red giant stars, and
compared our results to theoretical predictions based on the ``magnetic
greenhouse'' mechanism proposed by \citet{Fuller15} for dipole mode suppression.
Our results from modes of spherical degrees $\ell \leq 3$, provide
strong support for one of the main assertions of the theory behind the
mode suppression -- all mode energy leaking into the g-mode cavity of a
star is efficiently trapped or dissipated.  Specifically, we confirm the up
to almost 100\% suppression for dipole modes previously shown and
measure for the first time up to 49\% suppression on average of the
quadrupole modes and no suppression of the octupole modes.
We find it likely that a variation in the mode lifetime
along the red giant branch could explain the small difference in slope
between our prediction and the observations in the $V^2_{\ell=2}$ versus \numax\
diagram.  However, the variation required needs to be confirmed by
measurements of the mode lifetime of radial modes for a robust sample of
stars with \numax\ around 220\muhz\ and 70\muhz. 

Further tests of the magnetic greenhouse effect could be implemented with
determinations of the lifetimes of the suppressed dipole modes, or the
measurement of suppressed dipole mode visibilities in helium-burning clump
stars (Cantiello et al. 2016 in prep.) 

\begin{acknowledgements}
We acknowledge the entire {\it Kepler} team, whose efforts made these
results possible.  
D.S. is the recipient of an Australian Research Council
Future Fellowship (project number FT140100147). J.F. acknowledges support
from NSF under grant no. AST-1205732 and through a Lee DuBridge Fellowship
at Caltech.  R.A.G. acknowledge the support of the European Community{'}s
Seventh Framework Programme (FP7/2007-2013) under grant agreement
No. 269194 (IRSES/ASK), the ANR- IDEE (n ANR-12-BS05-0008), and from the
CNES.  D.H. acknowledges support by the Australian Research Council's
Discovery Projects funding scheme (project number DE140101364) and support
by the National Aeronautics and Space Administration under Grant NNX14AB92G 
issued through the Kepler Participating Scientist Program.
This project was supported by NASA under TCAN grant number NNX14AB53G, and
the NSF under grants PHY 11-25915 and AST 11-09174.  
Funding for the Stellar Astrophysics Centre is provided by The Danish
National Research Foundation (Grant agreement no.: DNRF106). The research
is supported by the ASTERISK project (ASTERoseismic Investigations with
SONG and Kepler) funded by the European Research Council (Grant agreement
no.: 267864). 
\end{acknowledgements}





\end{document}